\documentclass[sigconf,nonacm]{acmart}

\providecommand*{\Description}[1]{}
\usepackage{graphicx}
\usepackage{booktabs}
\usepackage{placeins}
\usepackage{algorithm}
\usepackage{array}
\usepackage{pifont}
\usepackage{subcaption}
\usepackage{multirow}
\usepackage[noend]{algpseudocode}
\usepackage[nameinlink,noabbrev]{cleveref}

\hypersetup{
  pdftitle={Retargetable Compilation for Tensor Kernels via a Unified Layout Abstraction},
  pdfauthor={Likai Ma; Renze Chen; Tianqi Li; Yuhao Luo; Youwei Xiao; Size Zheng; Xiuhong Li; Yun Liang}, 
  pdfsubject={},
  pdfkeywords={retargetable compilation, instruction layout, constraint solving}
}

\title{Retargetable Compilation for Tensor Kernels via a Unified Layout Abstraction}

\begin{document}

\begin{abstract}
Modern hardware accelerators achieve high performance through specialized
instructions coupled with tightly managed memory hierarchies. These
instructions impose strict layout requirements on their operands: where
tensor fragments reside, how they are organized, and how parallel units
access them. Such instruction-visible layout is critical to kernel
performance, as it determines whether instructions execute correctly and how
much data movement they incur.
However, existing compilers either scatter layout requirements across
handwritten programs and local scheduling rules, or provide layout
representations and conversions specific to a single class of targets.
Both approaches expose the same abstraction gap: the absence
of a unified model for instruction layout requirements across backends.

This paper presents \textbf{\textsc{Arknife}}, a retargetable compiler that treats
instruction-visible layout as first-class instruction semantics. Its central
idea is a constraint-based layout algebra: tensor layouts map logical
axes to hardware axes spanning parallel levels and memory spaces.
Instruction operands declare layout requirements, and target-provided repairs resolve 
conflicts. Layouts are then solved globally across the program.
ISA-specific layout facts are supplied by target-side instruction
descriptions rather than hardcoded in the compiler. Retargeting reuses
the layout algebra and solver while defining the new target's
instructions and their operand requirements.
We evaluate \textbf{\textsc{Arknife}} across x86 CPU, Hexagon NPU, and NVIDIA A100 GPU. 
\textbf{\textsc{Arknife}} matches cuBLAS on A100, and on
x86 and Hexagon outperforms vendor-optimized libraries by $2.17\times$ and
$3.78\times$ and the retargetable compiler MISAAL by $5.08\times$ and
$4.53\times$, respectively.
\end{abstract}

\author{Likai Ma}
\affiliation{%
  \institution{Peking University}
  \country{China}}

\author{Renze Chen}
\affiliation{%
  \institution{Peking University}
  \country{China}}

\author{Tianqi Li}
\affiliation{%
  \institution{Peking University}
  \country{China}}

\author{Yuhao Luo}
\affiliation{%
  \institution{Peking University}
  \country{China}}

\author{Youwei Xiao}
\affiliation{%
  \institution{Peking University}
  \country{China}}

\author{Size Zheng}
\affiliation{%
  \institution{Peking University}
  \country{China}}

\author{Xiuhong Li}
\affiliation{%
  \institution{Infinigence-AI}
  \country{China}}

\author{Yun Liang}
\affiliation{%
  \institution{Peking University}
  \country{China}}

\renewcommand{\shortauthors}{Likai Ma and Yun Liang, et al.}

\maketitle
\section{Introduction}

High-performance machine learning systems increasingly rely on specialized
hardware instructions such as tensor core MMA operations, AVX-512 VNNI, 
and Hexagon HMX matrix instructions~\cite{unit2021,
tensorir2022,amos2022}.  Extracting high throughput from such instructions 
requires coordinating specialized compute, local memories, explicit data movement, 
and hardware-specific parallel execution. This burden falls on both programmer and compiler.
For the programmer, hand-tuning is costly. A layout mistake produces wrong
results and manual optimization limits the design
space~\cite{cutlass,hao2026scaling}, while porting across hardware with different architectures means starting the whole effort over from scratch. 
A retargetable compiler can relieve this burden from programmer. 
However, compilers face the same structural obstacle. Arithmetic semantics alone 
do not capture an instruction's operand requirements. Correct execution also 
depends on where operands reside, how their dimensions are organized, and how 
their elements are distributed across parallel hardware units.

Layout is therefore part of the usable interface of an instruction, and the
challenge is to satisfy these instruction layout demands globally, so
that the data path is free of unnecessary movement and ad-hoc conversion.
A Hexagon kernel~\cite{hao2026scaling}, for example, mixes DMA transfers, 
HVX vector operations, and HMX matrix operations across three memory spaces,
each imposing a distinct operand layout. Stitching them into one data path
requires explicit layout conversion at the boundaries. The ISA provides over
a dozen dedicated primitives for this purpose alone, covering register
transposition (\texttt{vshuff}, \texttt{vdeal}), element-width conversion
(\texttt{vpack}, \texttt{vunpack}), and register reshaping (\texttt{vcombine},
\texttt{vsplat}, \texttt{valign}). We call instructions whose purpose is to
convert between layouts \emph{repairs}. A compiler that only checks
arithmetic equivalence may select the correct instructions yet still produce
an incorrect or repair-heavy data path.

These layout challenges are not confined to Hexagon. They also arise on CPU
and GPU, where matrix extensions and tensor instructions face the same
underlying problem~\cite{tensorir2022,tilelang2025,tirx2026}. In a GEMM
with epilogue, the tile matmul accumulates into MMA fragments, but the
epilogue reduction needs a row-major view for axis-aligned access. The
boundary between them requires explicit layout conversion. Hand-tuned
kernels such as FlashAttention-2's Ampere
implementation~\cite{flashattention22023} encode this conversion by hand.
A global layout solver can search for compatible layouts systematically
instead.

Existing systems address different aspects of the cross-backend layout
problem, but none provides a unified layout abstraction for retargetable compilation. 
Layout-guided tensor DSLs and scheduling 
systems~\cite{tensorir2022,tilelang2025,tirx2026} make data placement part 
of the program or schedule, so porting a kernel can require re-deriving 
layouts for each target. The impact of a layout rejected by an instruction goes beyond performance
to correctness. Synthesis-based
systems~\cite{hydride2024,misaal2025} move layout decision into the
compiler, but resolve mismatches locally, potentially leaving conversions that a 
global layout choice could eliminate. Constructive layout algebras
such as CuTe and Linear Layouts~\cite{cutlass,linear_layouts2026}
derive conversions between concrete layouts, but their conversions rely on GPU-specific 
ISA primitives for efficient lowering. On another ISA, the same
construction may have no efficient lowering, or a suitable conversion
algebra may have to be re-derived for the target. Generic fallbacks
such as scalarization can be orders of magnitude more expensive.
As a result, existing systems still struggle to combine consistent cross-backend 
programming with end-to-end performance.

To close this gap, we present \textbf{\textsc{Arknife}}, a
retargetable compiler for tensor kernels built around a layout
algebra that treats layouts as unknowns to be solved. We model each
target by its parallel hierarchy, memory spaces, and instructions,
with parallel indices and local addresses forming a hardware
coordinate space. A tensor layout is an injective mapping from
logical axis factors into this space, while an instruction operand 
declares a \emph{partial} mapping that constrains only the factors
the instruction observes. All requirements of a tensor's bound uses
therefore conjoin into a single accumulated constraint, and binding
legality reduces to its satisfiability rather than to layout
construction. The constraint system is order-independent, monotone,
and local to each instruction boundary, so the compiler can evaluate feasibility 
with a small integer program at any point of instruction search. 
An unsatisfiable constraint exposes the conflict, which a target-provided 
\emph{repair} resolves. A satisfiable one yields globally chosen layouts, 
eliminating conversions by layout choice rather than post-hoc synthesis.

Concretely, layout feasibility shapes every stage of
\textbf{\textsc{Arknife}} compilation, which falls naturally into two phases.
The first phase establishes layout feasibility.  For each operator,
\textbf{\textsc{Arknife}} enumerates the arithmetically equivalent target
instructions and commits the cheapest candidate whose accumulated constraint
remains satisfiable, with cost measured by the repairs that the candidate's
operand requirements would induce.  Wherever requirements still conflict
across a tensor's uses, it inserts the corresponding target-provided repairs,
so that every instruction boundary in the resulting graph is layout-feasible.
The second phase optimizes within this feasible space.  E-graph rewriting
first removes redundant repairs, after which a single integer program assigns
the remaining axis mappings, maximizing layout preferences while keeping every
instruction boundary legal.  The chosen layouts are then preserved through
tiling, memory planning, and code generation, down to the emitted kernel.

Together, the layout algebra and the compilation pipeline turn
layout compatibility from a programmer decision into a compiler
decision. \textbf{\textsc{Arknife}} reads operand requirements from target-side
instruction descriptions and checks them at instruction boundaries.
Retargeting requires defining the target's instructions through the
description interface, without changing the core layout algebra or
solver.

This paper makes the following contributions:

\begin{itemize}
  \item We introduce a target-independent layout algebra that expresses
  what each instruction demands of its operand layouts. Porting it to
  a new ISA requires no new algebraic transformations.

  \item We provide a consistent programming interface across
  backends, delegating instruction-visible layouts to the compiler.

  \item We build a retargetable compiler that removes redundant
  repairs through e-graph rewriting and performs global layout
  solving, delivering high-performance operators on heterogeneous
  backends.
\end{itemize}

We evaluate \textbf{\textsc{Arknife}} on x86 CPU, Qualcomm Hexagon NPU, and NVIDIA A100.
Compared with MISAAL, a state-of-the-art retargetable
compiler~\cite{misaal2025}, \textbf{\textsc{Arknife}} achieves $5.08\times$ and
$4.53\times$ geometric mean speedups on x86 and Hexagon benchmarks,
respectively. It also achieves $3.78\times$ over QNN~\cite{qnn}, the vendor
Hexagon library. On A100, it reaches $5.04\times$ over unfused
PyTorch~\cite{pytorch2019} on FP16 MHA and 97\% of cuBLAS~\cite{cublas} performance on TF32 GEMM.

\section{Background and Motivation}
\label{sec:background}

\subsection{Layout in Accelerator Compilation}
\label{sec:instruction-visible-layouts}

A tensor kernel depends on two classes of instructions: compute and conversion. 
Compute instructions perform arithmetic within one level of the memory hierarchy and assume their operands
already lie in a placement they accept: \texttt{\_mm512\_dpbusd\_epi32},
for instance, accumulates products of packed byte groups within a register.
Conversion instructions change that placement, either by moving data between levels
or by rearranging it within a level: \texttt{\_mm512\_unpacklo\_epi16} interleaves
the 16-bit elements of its two operands, and \texttt{vshuff} interleaves bytes within
a Hexagon vector register. A kernel is thus a chain of interleaved compute and 
conversion instructions, and either the programmer or the compiler must decide where the conversions
fall in that chain.

\begin{figure}[t]
  \centering
  \includegraphics[width=\columnwidth]{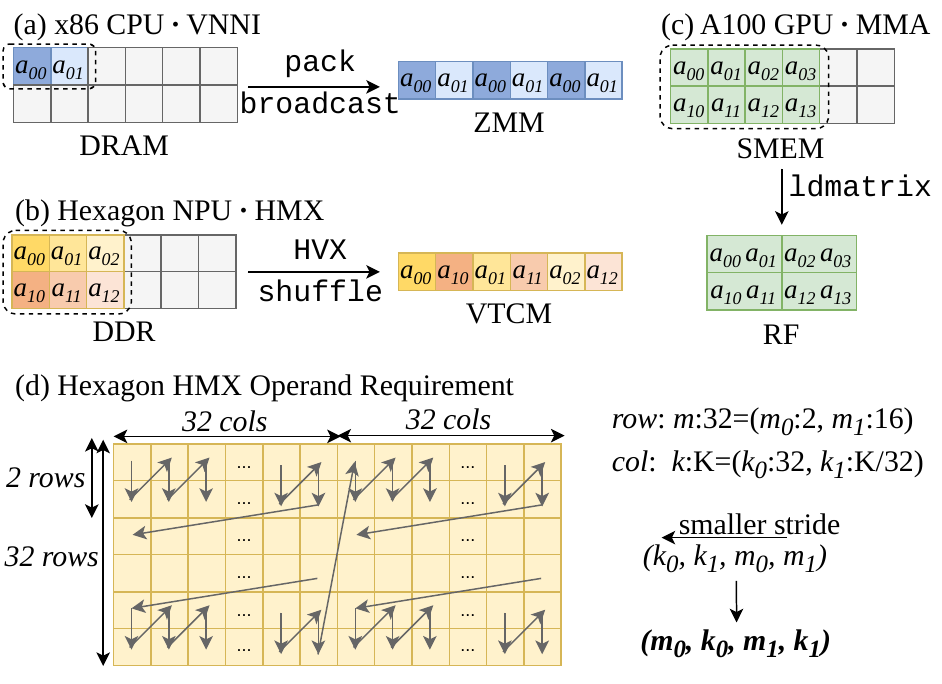}
  \caption{Three operand placements of the same logical GEMM $A$ tile,
  required by x86 VNNI, Hexagon HMX, and A100 MMA; (d) states the Hexagon
  requirement in full for two consecutive $32\times32$ FP16 blocks.}
  \Description{Four panels. (a) x86 VNNI: two adjacent elements of a row-major
  DRAM tile are packed and broadcast into one ZMM register. (b) Hexagon HMX:
  an HVX shuffle turns a row-major DDR tile into a VTCM row whose rows are
  interleaved in pairs. (c) A100 MMA: ldmatrix loads a shared-memory tile into
  per-lane registers. (d) Hexagon HMX requirement: a 32 by 32 FP16 tile in two
  consecutive 32-column blocks, rows interleaved in pairs, with a zig-zag
  tracing the order in which the fill reads the tile; beside it the row-major
  order of the four axis factors is set in plain type and the required order
  in bold.}
  \label{fig:retargetable-gemm}
\end{figure}

That decision is hard, and the requirements differ from one backend to
another. \Cref{fig:retargetable-gemm}(a)--(c) show three placements of the same logical GEMM
$A$ tile. On x86, adjacent elements are packed into one ZMM register; 
on Hexagon, rows are interleaved in pairs in VTCM; and on A100, 
\texttt{ldmatrix} distributes elements across lanes and registers. 
The consuming instruction imposes each layout and states it in its vocabulary. 
We call such a requirement an
\emph{instruction-visible layout}. Although absent from the instruction's arithmetic, 
it determines whether the operand is legal.

Such a requirement may be finer than whole-axis ordering.
\Cref{fig:retargetable-gemm} gives the order in full for a $32\times32$ FP16
tile consumed by HMX, from innermost to outermost: the two rows of each
pair, the 32 columns of a block, the 16 row pairs, and the next
reduction-axis block. Because the order mixes factors of both axes,
whole-axis permutations such as row-major or column-major cannot express it.
The HMX tile-buffer fill follows this order, so any other arrangement is an
invalid operand, not merely a slower one. This requirement reflects a broader
difficulty: \textbf{cross-backend layout handling is central to retargetable
tensor-kernel generation}. Instruction-visible layouts are imposed at
different memory and parallel levels, in different vocabularies. A layout
legal on one target may be invalid on another, and a conversion cheap on one
may be prohibitively expensive on another.

\subsection{Motivation}

\begin{figure}[t]
  \centering
  \includegraphics[width=\columnwidth]{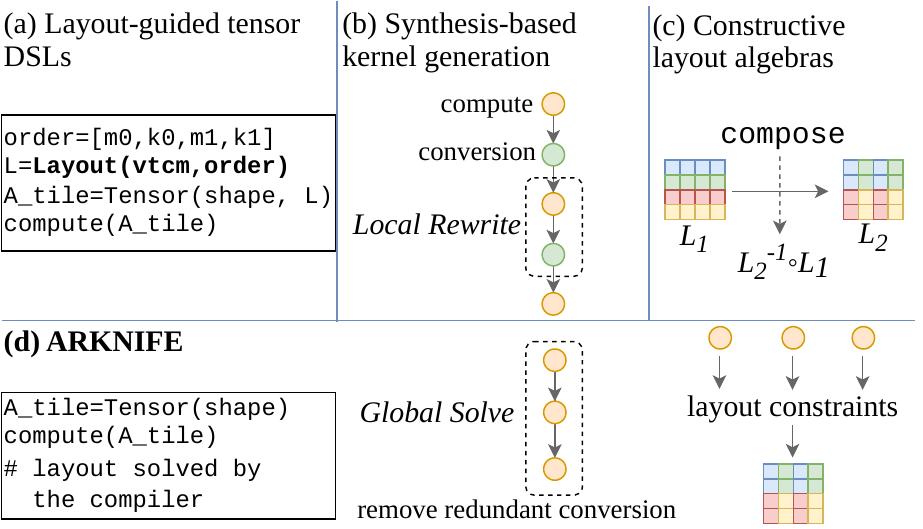}
  \caption{Layout handling in existing systems (a)--(c) and in
  \textbf{\textsc{Arknife}} (d). The dashed box marks the scope of
  mismatch resolution: the region matched by a rewrite in (b), the whole
  graph in (d).}
  \Description{Four panels. (a) A program constructs a tensor from an explicit
  memory space and axis order before calling compute. (b) A chain alternates
  compute and conversion nodes, with a dashed box around a short segment
  labelled local rewrite. (c) Two concrete layouts L1 and L2 are joined by an
  arrow labelled compose. (d) spans the width under the other three: the same
  program without the layout argument, commented that the layout is solved by
  the compiler; the same chain with a dashed box around the whole chain
  labelled global solve, annotated that redundant conversions are removed; and
  three compute nodes whose layout constraints feed one solved layout.}
  \label{fig:contrast}
\end{figure}

Systems address this layout-handling difficulty differently.
\Cref{fig:contrast} contrasts three current approaches (a)--(c) with the
design of \textbf{\textsc{Arknife}} (d) along three dimensions: who states
the layout, the scope of mismatch resolution, and the layout algebra. 

In layout-guided tensor DSLs and schedules, the layout starts as something the
program states. TensorIR~\cite{tensorir2022} fixes it in the schedule, and
TIRx~\cite{tirx2026} attaches it to the tensor type, where the user declares tile
placement across lanes, warps, and registers. For a single target this is merely
verbose, while across backends it becomes a heavy burden, as the user must
re-derive the placement for each target's hierarchy and instruction constraints.
Worse, an unlucky or unchecked layout choice surfaces as incorrect results, not
merely lost performance.
\textbf{\textsc{Arknife}} aims to eliminate this burden
(\Cref{fig:contrast}(d)). The program drops the layout argument and the
compiler derives the placement.

Synthesis-based approaches take the layout decision into the compiler but
keep mismatch resolution local (\Cref{fig:contrast}(b)), with each rewrite
patching mismatches in isolation and so potentially missing conversions that a
global layout choice could eliminate. Hydride
and MISAAL~\cite{hydride2024,misaal2025}, for instance, take a target-scheduled IR as
input. Hydride synthesizes small windows of an
expression separately, choosing each conversion from local information even
when a layout choice one window away would have made it unnecessary. Solving
globally across the graph, as in \Cref{fig:contrast}(d), avoids this.
Concretely, compared to \textbf{\textsc{Arknife}}, Hydride executes
$17.1\times$ and $51.4\times$ as many layout-handling instructions on the two
x86 fused graphs, and $85.8\times$ as many on Hexagon quantized Add
(\Cref{sec:layout-handling-breakdown}).

Constructive layout algebras, \Cref{fig:contrast}(c), take layouts as
input. Composing two concrete layouts yields the conversion between
them. CuTe~\cite{cutlass} composes and divides layouts this way and Linear
Layouts~\cite{linear_layouts2026} does the same over bit-level mappings.
In both, the layout must already be known before the algebra runs. This
suffices where layout-change mechanisms are few, fixed in the ISA, and
comparably priced. Within a GPU block, for instance, a shared-memory round
trip converts any layout into any other at bounded cost. A retargetable compiler has
none of these constants. In \Cref{fig:retargetable-gemm}(d), an HMX tile
in VTCM has rows interleaved in pairs and its vector consumer expects
row-major registers. The permutation between them crosses the
VTCM--register boundary, where no comparably priced primitive
exists. The fallback is byte-granular scalar moves at 70--504 cycles
per 128 bytes, against 3.5 in-register.\footnote{Cycle counts from
\texttt{hexagon-sim} (\texttt{-mv66}, HVX-128B, single thread,
issue-bound dependent chains). The universal in-register permute is the
\texttt{vrdelta}/\texttt{vdelta} Benes pair (Hexagon V75 HVX
Programmer's Reference Manual, \S6.10).} 
This motivates a constraint-based algebra that operates on requirements
rather than layouts (\Cref{fig:contrast}(d)). Each instruction
contributes a partial operand requirement, and one layout is solved from
all of them. The algebra itself would remain fixed across backends, and only
its inputs would change, making the formulation retargetable. The two
formulations answer different questions---given layouts, derive the
transformation or given requirements, decide whether a layout exists and
which to choose.

Across all three approaches, what an instruction requires of its operands 
remains implicit, woven into program text, local rewrite rules, or 
target-dependent layout compositions.
Retargeting re-encodes compatibility rules from scratch.
\textbf{\textsc{Arknife}} therefore pursues two design goals. The first is
representational: one layout algebra, tied to no vendor ISA, that decides
whether an operand is legal on any target (\Cref{sec:target-abstraction}).
The second is algorithmic: remove redundant repairs and solve
layouts globally across the compute graph (\Cref{sec:compilation}).
\section{Overview}
\label{sec:overview}

\begin{figure}[t]
  \centering
  \includegraphics[width=\linewidth]{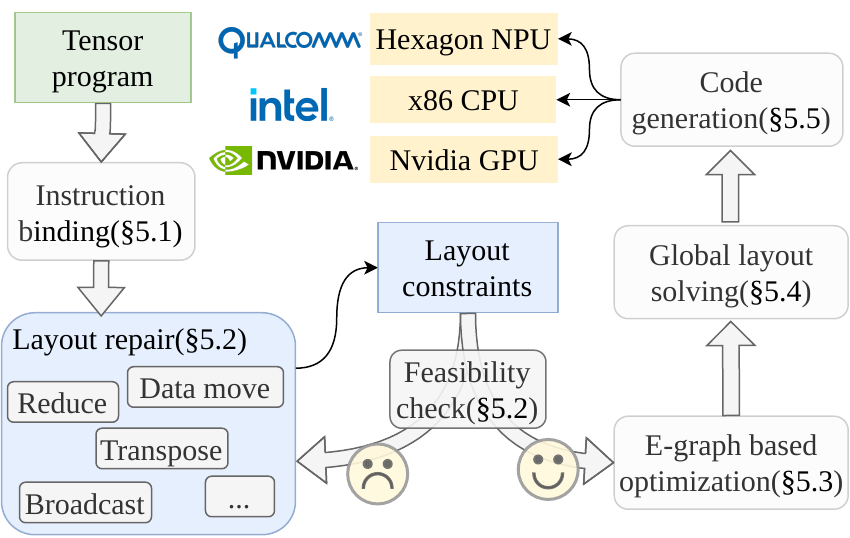}
  \caption{Overview of the \textbf{\textsc{Arknife}} compilation workflow.}
  \Description{A flow diagram with tensor program and target description inputs,
  instruction binding, layout repair choices, layout constraints, feasibility
  checking, e-graph optimization, global layout solving, and code generation.}
  \label{fig:system-overview}
\end{figure}

\Cref{fig:system-overview} shows the \textbf{\textsc{Arknife}} compilation
workflow. The compiler takes a tensor program and a target description as
input. The program specifies the logical computation, while the target
description provides parallel levels, memory spaces, instruction templates,
and their operand-layout requirements. The output is an executable kernel
specialized to the target. \Cref{fig:compile-example} instantiates the
pipeline on a quantized-add excerpt for the Hexagon target.

Compilation begins with instruction binding (\Cref{sec:binding}), which
considers candidate target instructions for each program operation. For
each candidate, the compiler checks whether its operand-layout
requirements can be satisfied directly or through target-provided
repairs, such as \emph{reduce}, \emph{data move}, \emph{transpose}, and
\emph{broadcast}. The layout algebra (\Cref{sec:target-abstraction})
represents tensor layouts and operand requirements as mappings from
logical axes to hardware axes. Repair insertion and feasibility checking
(\Cref{sec:repair}) then rejects combinations of candidate bindings and
repairs whose axis-level constraints cannot be satisfied jointly. An
e-graph rewriting pass (\Cref{sec:repair-opt}) simplifies the repair graph
by eliminating redundant conversions. A global solver
(\Cref{sec:global-solving}) assigns concrete tensor layouts subject to the
remaining constraints. Finally, layout-preserving lowering and code
generation (\Cref{sec:lowering}) emit the executable kernel.

\section{A Layout Algebra for Tensor Programs}
\label{sec:target-abstraction}

This section develops the layout algebra that makes instruction binding
a layout-aware decision. \Cref{sec:target-model} models the target as
$\mathcal{T}=(L,M,I,R)$ and defines the hardware coordinate space $H$.
\Cref{sec:coordinate-layout} formalizes logical spaces and layouts. 
\Cref{sec:axis-layout} introduces
the factorized axis-level representation used for solving, and
\Cref{sec:layout-compatibility} reduces boundary legality to a least
constraint on the tensor layout and repair to a small two-tensor
feasibility check. \Cref{tab:layout-notation} summarizes the notation.

\subsection{Target Model}
\label{sec:target-model}

To express the instruction-visible layouts of \Cref{sec:background}
uniformly, we model a target without fixing its ISA or the number of
parallel levels.

\begin{definition}[Target model]
\label{def:target-model}
A target is a tuple $\mathcal{T}=(L,M,I,R)$ of a finite set $L$ of
parallel levels, a finite set $M$ of memory spaces, a finite set $I$
of instruction templates, and the subset $R\subseteq I$ of repair
instructions, which change a tensor's representation through memory
movement, layout permutation, or type conversion.
\end{definition}

\begin{definition}[Hardware coordinate space]
\label{def:hardware-space}
Each parallel level $l$ of $n_l$ units contributes a parallel-coordinate axis
$S_L(l)$ of extent $n_l$, and each memory space $m$ contributes a per-unit
local address space $S_M(m)$. Their Cartesian product defines the hardware
coordinate space
\[
H=\Bigl(\prod_{l\in L}S_L(l)\Bigr)\times\Bigl(\prod_{m\in M}S_M(m)\Bigr),
\]
whose axes are called \emph{hardware axes}, parallel-coordinate axes
contributed by $L$ and local-address axes contributed by $M$.
\end{definition}

\begin{figure*}[t]
  \centering
  \includegraphics[width=\textwidth]
    {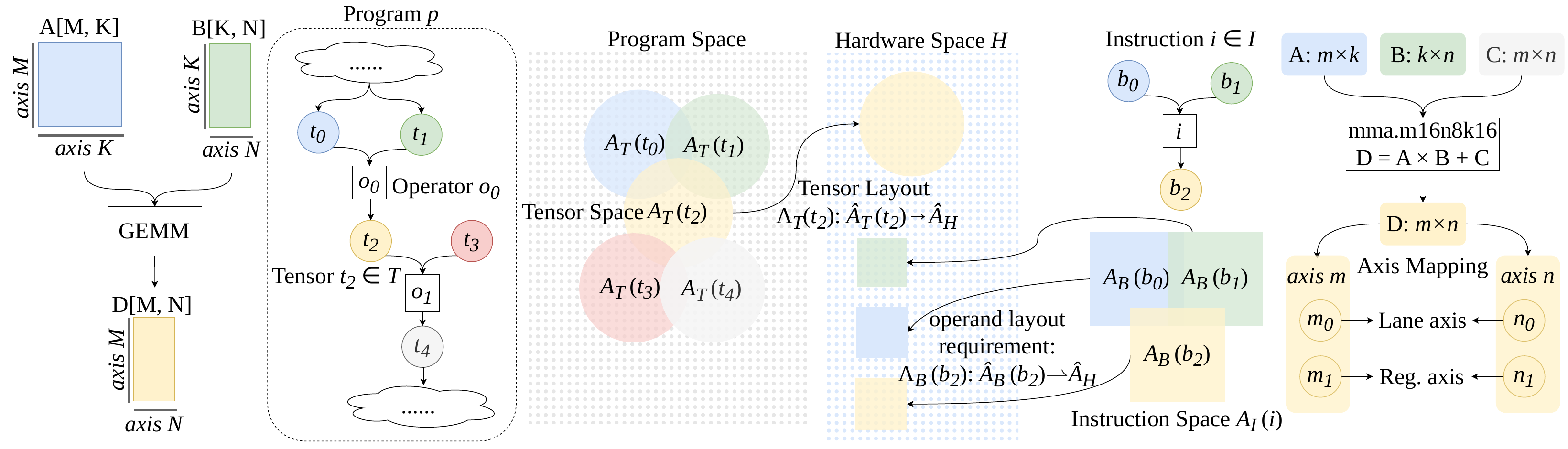}
  \caption{Program and instruction spaces map into $H$.}
  \Description{A GEMM program instance and an MMA instruction instance
  illustrate how program tensors and instruction operands carry logical
  spaces that map into the hardware memory space $H$.}
  \label{fig:space-mapping}
\end{figure*}

\begin{table}[t]
  \caption{Notation in the layout algebra.}
  \Description{Frequently used symbols for hardware space, logical spaces, layouts,
  operand views, and boundary constraints.}
  \label{tab:layout-notation}
  \centering
  \footnotesize
  \begin{tabular}{
    @{}
    >{\raggedright\arraybackslash}p{0.20\columnwidth}
    >{\raggedright\arraybackslash}p{0.77\columnwidth}
    @{}
  }
    \toprule
    \textbf{Symbol} & \textbf{Meaning} \\
    \midrule
    $H$
      & Hardware coordinate space. \\

    $S_T(t)$, $S_B(b)$
      & Logical index spaces of tensor $t$ and instruction operand $b$. \\

    $A_T(t)$, $A_B(b)$
      & Logical axis sets of tensor $t$ and instruction operand $b$. \\

    $\widehat{A}_T(t)$, $\widehat{A}_B(b)$, $\widehat{A}_H$
      & Factor sets obtained by factoring the tensor, operand, and hardware
        axes, respectively. \\

    $F_T(t)$, $F_B(b)$
      & Tensor layout and operand constraints in hardware coordinates. \\

    $\Lambda_T(t)$, $\Lambda_B(b)$
      & Factor mappings representing tensor layouts and operand
        requirements, respectively. \\

    $\phi_{o,t,b}$
        & Embedding of operand factors into tensor factors. \\
    
    $V_{o,t,b}$
        & Tensor layout viewed by the operand:
          $\Lambda_T(t)\circ\phi_{o,t,b}$. \\

    $B_{OT}(o,t)$
        & Operand corresponding to $t$ in the instruction bound to $o$. \\

    $U_t$
      & Bound operand uses of $t$, including producers and consumers. \\

    $C_t$
      & Union of operand requirements mapped to tensor factors. \\
    \bottomrule
  \end{tabular}
\end{table}

\subsection{Coordinate-Level Layout Problem}
\label{sec:coordinate-layout}

Tensor programs and hardware instructions are indexed by multidimensional
integer domains, whose independent dimensions we call \emph{axes}.

\begin{definition}[Axes and logical spaces]
\label{def:logical-space}
An axis $a$ is a one-dimensional integer coordinate domain
with extent $r_a=|a|>1$. A logical space is the Cartesian product of a
finite collection $A$ of independent axes, written
$S=\prod_{a\in A}a$, with $\operatorname{axes}(S)=A$. Tensor logical
spaces in this work are domains of this form. Ragged domains and
data-dependent access relations are outside the model.
\end{definition}

Each tensor $t$ and instruction operand $b$ has a logical space $S_T(t)$
and $S_B(b)$, respectively. For example, the GEMM tensors $A[M,K]$,
$B[K,N]$, and $D[M,N]$ have $S_T(A)=M\times K$,
$S_T(B)=K\times N$, and $S_T(D)=M\times N$. A point
$(m,k)\in S_T(A)$ identifies an element of $A$.

\begin{definition}[Tensor layouts and operand requirements]
\label{def:coordinate-layouts}
A \emph{tensor layout} for a resident tensor tile $t$ is an injective
coordinate map $F_T(t):S_T(t)\rightarrow H$ that assigns each tensor
element a hardware location.

An \emph{operand layout requirement} for an instruction operand $b$
is a pair $F_B(b)=(\pi_b,g_b)$, where
$\pi_b:H\rightarrow H_b$ projects a hardware location onto the
coordinate components constrained by the instruction, and
$g_b:S_B(b)\rightarrow H_b$ specifies their required values.
An operand placement $F:S_B(b)\rightarrow H$ satisfies this
requirement exactly when $\pi_b\circ F=g_b$.
The remaining hardware-coordinate components are not constrained
by this requirement.
\end{definition}

For example, consider the two-vector interleaving illustrated in
\Cref{fig:layout-lcr}(a), where even- and odd-indexed elements occupy
separate vectors.
Writing a hardware location as $(q,p,v)$
(vector pair, vector within the pair, and element position),
its requirement is
  $\pi_b(q,p,v)=(p,v),
  g_b(k)=(k\bmod 2,\lfloor k/2\rfloor)$.
Thus, element $k$ must occupy position $\lfloor k/2\rfloor$ in
vector $k\bmod 2$, while the allocated pair $q$ remains unspecified.
The layout-mapping problem is to choose complete tensor layouts
that satisfy all bound operand requirements.

\subsection{Factorized Axis-Level Representation}
\label{sec:axis-layout}

Coordinate-map search assigns a hardware coordinate to every logical
point, yielding a search space exponential in tensor size. Therefore,
\textbf{\textsc{Arknife}} exploits the regularity of tensor layouts by
solving mappings between logical and hardware axes rather than between
individual points.

We write $A_T(t)=\operatorname{axes}(S_T(t))$ and
$A_B(b)=\operatorname{axes}(S_B(b))$ for the tensor and operand axis
sets. In the GEMM example, $A_T(A)=\{M,K\}$ and
$A_T(B)=\{K,N\}$, so their shared reduction axis is captured by
$A_T(A)\cap A_T(B)=\{K\}$.
Any axis may be factored as an ordered product 
$a\cong a_0\times\cdots\times a_q$, where
$r_a=\prod_{j=0}^{q}r_{a_j}$. Axis factorization enables fine-grained mappings beyond whole-axis
permutations. For example, it can express the HMX operand layout in
\Cref{fig:retargetable-gemm}(d) by independently mapping factors of
different logical axes and interleaving them in hardware order. Factoring the
axes in $A_T(t)$ and $A_B(b)$, together with the hardware axes of $H$,
yields the tensor, operand, and hardware factor sets
$\widehat{A}_T(t)$, $\widehat{A}_B(b)$, and $\widehat{A}_H$,
respectively.
A tensor layout is represented by an injective factor mapping
$\Lambda_T(t):\widehat{A}_T(t)\hookrightarrow\widehat{A}_H$,
while an operand requirement is represented by a partial factor mapping
$\Lambda_B(b):\widehat{A}_B(b)\rightharpoonup\widehat{A}_H$.
Mapped factors have equal extents and matching ordinal coordinates.
Each assignment $\Lambda_B(b)(a)=h$ requires $F(s)_h=s_a$ for every
operand point $s$. Collectively, these equalities define
$\pi_b\circ F=g_b$, leaving other hardware components unconstrained.
In \Cref{fig:space-mapping}, $\Lambda_T(t_2)$ maps $K$ and $N$ factors
to parallel and local-address axes, while the MMA requirement
$\Lambda_B(b)$ maps $m_0,n_0$ to lane axis factors and $m_1,n_1$ to register
axis factors.

This factorization restricts \(F\) to mappings decomposable across
axis factors, making layout constraints separable for global solving.
The core algebra models instruction-visible layout requirements that 
determine operand legality. Cross-axis optimizations that do not alter 
an instruction’s required operand view, such as shared-memory XOR swizzling 
for bank-conflict avoidance, are selected by the backend after layout solving 
and must preserve the solved instruction-boundary mappings.

\subsection{Layout Compatibility and Repair}
\label{sec:layout-compatibility}

\begin{figure}[t]
  \centering
  \includegraphics[width=\columnwidth]
    {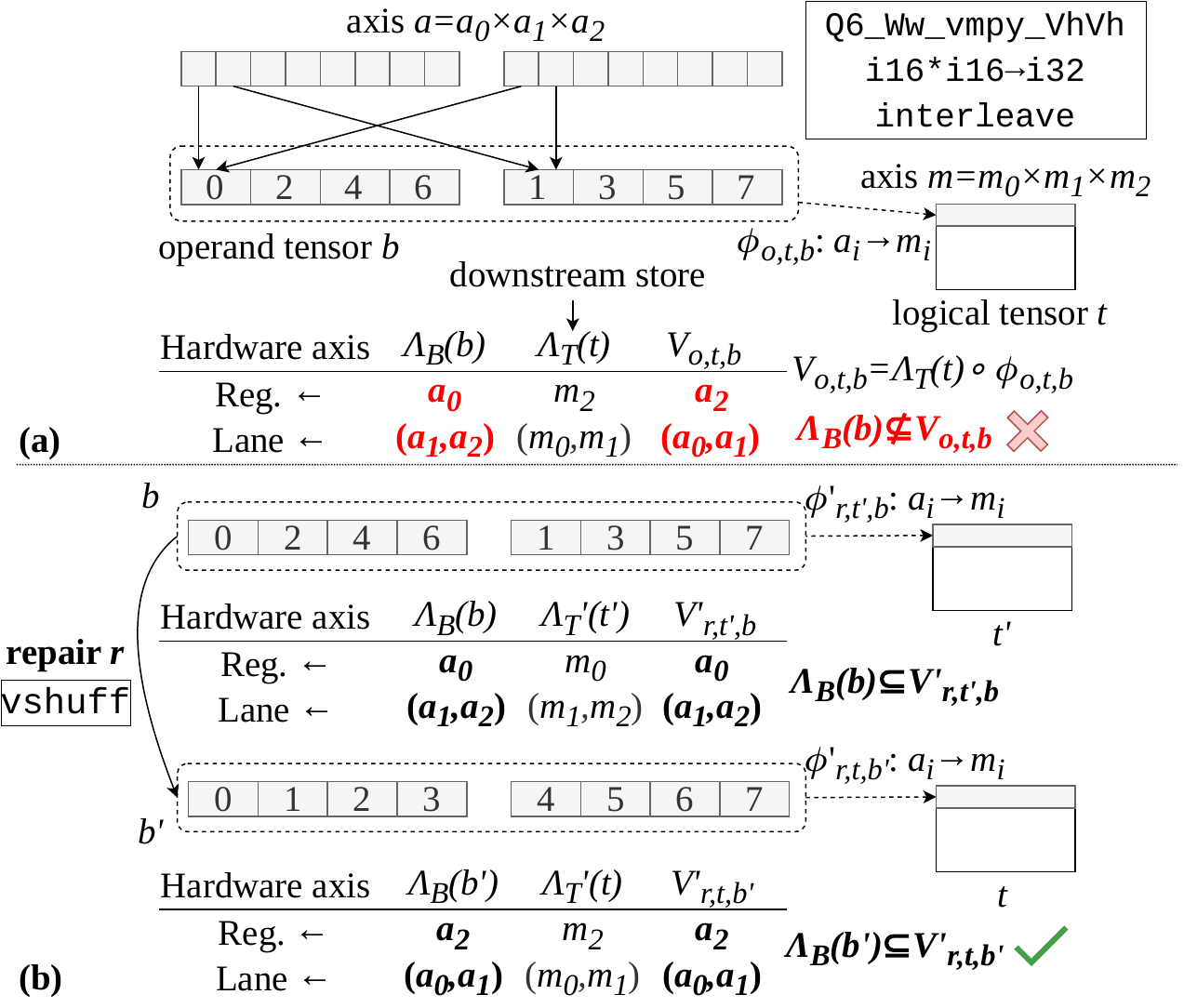}
  \caption{Layout conflict (a) and repair (b).
  All factors have extent two.}
  \Description{Two panels illustrate an axis-factor layout conflict and its
  repair. In panel (a), the operand and tensor mappings assign corresponding
  factors to different lane and register axes. In panel (b), repair r inserts
  an intermediate tensor t prime so that the axis-factor mappings agree at
  both resulting boundaries.}
  \label{fig:layout-lcr}
\end{figure}

Since axis-factor mappings induce their coordinate-level counterparts,
  boundary agreement can be checked directly at the axis level.
\Cref{fig:layout-lcr} illustrates a concrete conflict and its repair.

\begin{definition}[Operand view and legality]
\label[definition]{def:operand-legality}
Suppose operand $b$ of the instruction bound to operator $o$ uses tensor
$t$. Let $\phi_{o,t,b}:\widehat{A}_B(b)\hookrightarrow\widehat{A}_T(t)$ 
be an extent-compatible embedding that identifies, for each
instruction-local operand factor, the tensor factor it denotes. Thus,
$\phi_{o,t,b}$ describes a logical operand--tensor correspondence,
rather than a mapping into hardware.

The induced operand view must agree with $\Lambda_B(b)$ on its domain:
\[
  \begin{aligned}
    V_{o,t,b}
      &= \Lambda_T(t)\circ\phi_{o,t,b}
      : \widehat{A}_B(b)\rightarrow\widehat{A}_H, \\
    \Lambda_B(b)\subseteq V_{o,t,b}
      &\iff
      \forall a\in\operatorname{dom}(\Lambda_B(b)),\,
      V_{o,t,b}(a)=\Lambda_B(b)(a).
  \end{aligned}
\]
The bound operand is legal, written $\operatorname{Legal}(o,t,b)$, when
$t$ resides in the memory space required by $b$ and this agreement holds.
\end{definition}

In \Cref{fig:layout-lcr}(a), consider the candidate embedding
$\phi_{o,t,b}(a_i)=m_i$, which pairs each operand factor with its
corresponding tensor factor. The operand therefore inherits the
tensor's hardware mapping:
$V_{o,t,b}(a_i)=\Lambda_T(t)(m_i)$. For the shown tensor layout, this
view places $a_2$ on a register-axis factor and $a_0,a_1$ on lane-axis
factors, whereas $\Lambda_B(b)$ requires $a_0$ on a register-axis
factor and $a_1,a_2$ on lane-axis factors. Thus,
$\Lambda_B(b)\nsubseteq V_{o,t,b}$, and the direct binding is illegal.

\begin{definition}[Shared-layout compatibility]
\label{def:shared-compatibility}
Let $U_t$ be the set of bound uses of $t$. Index its $k$ elements as
$(o_j,b_j)$, where $b_j=B_{OT}(o_j,t)$. The uses are compatible when
there exist a valid factor mapping $\Lambda_T(t)$ and embeddings
$\{\phi_{o_j,t,b_j}\}_{j=1}^{k}$ such that every use is legal:
\[
  \operatorname{Compat}(t,U_t)
  \Longleftrightarrow
  \exists \Lambda_T(t),\{\phi_{o_j,t,b_j}\}_{j=1}^{k}:
  \bigwedge_{j=1}^{k}
  \operatorname{Legal}(o_j,t,b_j).
\]
\end{definition}

For fixed resource-valid bindings and embeddings, write
$\phi_j=\phi_{o_j,t,b_j}$ and
$C_t=\bigcup_{j=1}^{k}(\Lambda_B(b_j)\circ\phi_j^{-1})$.

\begin{proposition}[Least boundary constraint]
\label[proposition]{prop:least-boundary-constraint}
A valid layout $\Lambda_T(t)$ satisfies every use in $U_t$ exactly when 
$C_t\subseteq\Lambda_T(t)$.
\end{proposition}

\begin{proof}
Since each $\phi_j$ is injective, $\phi_j^{-1}$ is well-defined on its
image. By \Cref{def:operand-legality}, use $(o_j,b_j)$ is legal exactly
when $\Lambda_B(b_j)\circ\phi_j^{-1}\subseteq \Lambda_T(t)$. 
All uses are therefore legal exactly when
$C_t\subseteq \Lambda_T(t)$, and the union construction makes $C_t$ the least such
constraint.
\end{proof}

As a union, $C_t$ is independent of the order in which the uses in
$U_t$ are visited, and for a fixed binding it is a pure function of
the bound graph. It also grows monotonically with the use set:

\begin{proposition}[Monotonicity]
\label[proposition]{prop:monotonicity}
Writing $C_t(U)$ for the constraint induced by a use set $U$, if
$U\subseteq U'$ then $C_t(U)\subseteq C_t(U')$ and
$\operatorname{Compat}(t,U')\Rightarrow\operatorname{Compat}(t,U)$.
\end{proposition}

\begin{proof}
By \Cref{prop:least-boundary-constraint}, all uses in $U$ are legal
exactly when $C_t(U)\subseteq\Lambda_T(t)$. Union is monotone, and
any layout satisfying $C_t(U')$ satisfies $C_t(U)$.
\end{proof}

\begin{definition}[Layout-feasible repair]
\label[definition]{def:feasible-repair}
For a repair $r$ inserted after a producer use $u=(o,b)\in U_t$, let
$t'$ be produced at $u$ and consumed by $r$, whose output is $t$.
Let $b_r^{\mathrm{in}}$ and $b_r^{\mathrm{out}}$ be the input and output
operands of $r$, and define
$U_{r,u}^{t}=(U_t\setminus\{u\})\cup\{(r,b_r^{\mathrm{out}})\}$ and
$U_{r,u}^{t'}=\{u,(r,b_r^{\mathrm{in}})\}$. Then
\[
  \operatorname{Feasible}(r,t,U_t,u)
  \Longleftrightarrow
  \operatorname{Compat}(t,U_{r,u}^{t})
  \land
  \operatorname{Compat}(t',U_{r,u}^{t'}).
\]
\end{definition}

A target may provide a parameterized repair family, such as
$\{\mathit{Perm}_p\}$.  Given incompatible requirements at a boundary,
the compiler instantiates or enumerates family members whose input and
output constraints cover the two sides of the conflict.
\Cref{def:feasible-repair} then checks whether an instance separates the
conflicting requirements into two compatible tensors. 
In \Cref{fig:layout-lcr}(b), $b$ and $b'$ denote the repair
operands $b_r^{\mathrm{in}}$ and $b_r^{\mathrm{out}}$.
Repair $r$ converts producer-compatible $t'$ into
store-compatible $t$, 
preserving the tensor’s logical value semantics while allowing 
$t'$ and $t$ to have different layouts, and making both instruction boundaries compatible.
Such repairs include shuffles between lane and register layouts
and copies across memory spaces.

By \Cref{def:feasible-repair}, $\operatorname{Feasible}(r,t,U_t,u)$
involves only the tensors $t$ and $t'$. Each feasibility
check in repair selection (\Cref{sec:repair}) is therefore a small
model over the axes of two tensors, not the program graph
(\Cref{sec:global-solving}), which keeps the per-candidate checks
of the binding search affordable while enabling sound pruning 
during binding search (\Cref{prop:monotonicity}). The measured
overhead appears in \Cref{sec:layout-solving-overhead}.
\section{Layout-Aware Compilation}
\label{sec:compilation}

Given a target $\mathcal{T}=(L,M,I,R)$ and a program graph
$G_P$ with operator set $O$, the compiler first binds each operator to a
target instruction template, yielding a complete binding $b:O\to I$.
The selected instructions and any inserted repairs form a bound instruction
graph $G_I$, for which the compiler computes a layout assignment $\Lambda$.
A valid result requires
$\operatorname{Legal}$ at every bound operand,
$\operatorname{Compat}$ at every shared tensor, with feasible repairs
from $R$ inserted as needed, and every temporary to fit its assigned memory
region.

\begin{figure*}[t]
  \centering
  \includegraphics[width=\textwidth]{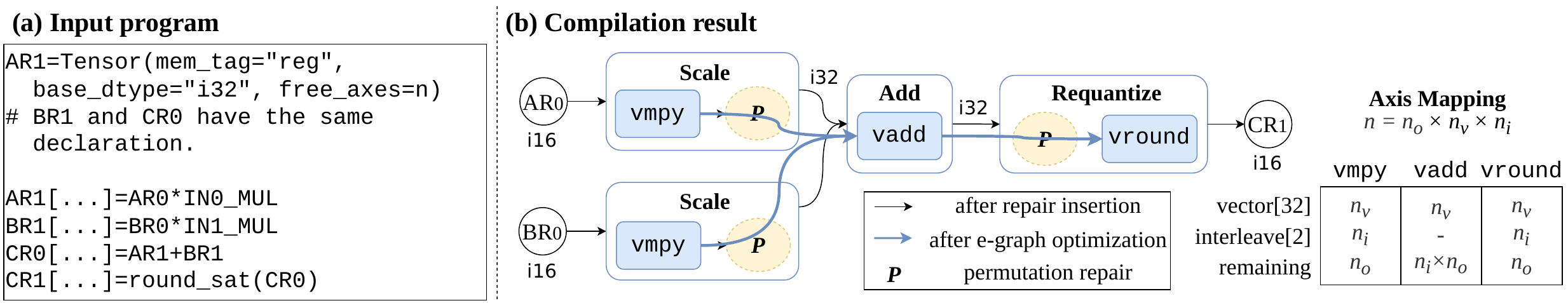}
  \caption{Compilation of quantized addition on Hexagon.}
  \Description{Two panels illustrate quantized-add compilation.
  Panel (a) declares an i32 register tensor with free axis n and shows
  two multiplications, an addition, and requantization.
  Panel (b) maps the multiplications to vmpy, addition to vadd, and
  requantization to vround, with i16 inputs and output and i32 intermediates.
  Black paths contain three permutation repairs marked P; blue paths
  bypass them after e-graph optimization.
  The mapping table decomposes n into n_o, n_v, and n_i.
  All three instructions share vector factor n_v; vmpy and vround also
  use interleave factor n_i. Remaining axes are n_o for vmpy and vround,
  and n_i times n_o for vadd; these lie outside the respective instruction tiles.}
  \label{fig:compile-example}
\end{figure*}

\Cref{fig:compile-example}(a) shows a Hexagon quantized-add excerpt,
omitting normalization and clamping. The program declares register tensors
with free logical axes and instruction-level axis mappings remain unspecified.
\Cref{fig:compile-example}(b) shows instruction binding, repair simplification, and solved mappings.

\subsection{Instruction Binding}
\label{sec:binding}

\noindent\textbf{\textit{Binding problem.}}
Instruction binding maps each program operator to a target instruction
template with equivalent arithmetic semantics. For an operator $o$, the
candidates are
\[
  \operatorname{Cand}(o)
  =
  \left\{
    \iota \in I
    \mid
    \operatorname{Semantics}(\iota) \models o
  \right\},
\]
where $\models$ denotes arithmetic equivalence. During candidate selection, a binding
$b:O\rightharpoonup I$ is complete when it covers every operator and
layout-feasible when the operand requirements of the selected instructions
can be satisfied, possibly through repairs.

\noindent\textbf{\textit{Candidate selection.}}
The compiler visits operators in topological order and ranks candidate
instructions using a heuristic that favors small register footprints and
compatibility with already-bound neighbors. It checks each candidate against
the accumulated layout constraints, rejecting infeasible choices and
backtracking as needed.

In \Cref{fig:compile-example}(b), Scale, Add, and Requantize bind
to \texttt{vmpy}, \texttt{vadd}, and \texttt{vround}, respectively.
The layout required by \texttt{vmpy} distributes 64 products across two
32-element \texttt{i32} vectors, and \texttt{vround} consumes this
organization to produce \texttt{i16} values.
\texttt{vadd} requires only a common 32-element vector
axis across its operands. These requirements specify local layouts
but leave their correspondence to program axis $n$ unresolved.

\subsection{Repair Insertion and Feasibility Checking}
\label{sec:repair}

\noindent\textbf{\textit{Repair insertion.}}
Instruction binding introduces instruction-local axes and their operand
requirements.  Under the current local axis correspondences, a shared
tensor boundary may induce incompatible layout requirements.  The
compiler makes such a mismatch explicit by inserting the lowest-cost
target-provided repair instance satisfying
\Cref{def:feasible-repair}:
\[
  r^\star=\operatorname*{arg\,min}_{r\in R}\operatorname{Cost}(r)
  \;\text{s.t.}\;
  \operatorname{Feasible}(r,t,U_t,u).
\]

These repairs are provisional because the following e-graph pass may move,
compose, or eliminate them before global axis-map solving assigns the
remaining free mappings.
In \Cref{fig:compile-example}(b), initial binding handles
instruction-local layouts separately. Two permutation repairs
reorder the \texttt{vmpy} outputs, while a third restores the
organization consumed by \texttt{vround}.
Yet \texttt{vadd} can operate independently on the two vectors,
allowing the interleaved representation to persist across the addition.

\noindent\textbf{\textit{Incremental feasibility checking.}}
Before committing a tentative instruction binding or repair, the
compiler checks feasibility of the affected subgraph's axis constraints
and rejects infeasible candidates, soundly by \Cref{prop:monotonicity}.
Instructions without axis constraints, such as scalar loads and plain
copies, skip this check. In the running example, all three permutation
repairs undergo this check, and their axis constraints enter the global
layout problem.

\subsection{Repair Graph Optimization}
\label{sec:repair-opt}

\textbf{\textsc{Arknife}} removes redundant transforms from repair-heavy
dataflow slices using e-graph saturation and cost-based
extraction~\cite{egg2021}.
Let $\mathit{Perm}_p$ denote a tensor permutation
that applies componentwise to a tensor tuple
$\mathbf{A}=(A_1,\ldots,A_m)$, and let $\mathit{Bcast}$ denote a broadcast. 
\Cref{tab:repair-rewrites} summarizes four
representative generic rewrites, and backends may add target-specific rules.

\begin{table}[t]
  \centering
  \caption{Representative repair-graph rewrite rules.}
  \Description{Four e-graph rewrite rules, Commute, Fusion, Identity, and
  Hoist, are listed with their symbolic transformations. Additional
  applicability conditions are shown for Commute and Hoist.}
  \label{tab:repair-rewrites}
  \small
  \begin{tabular}{
    l
    >{\raggedright\arraybackslash}p{0.69\columnwidth}
  }
    \hline
    \textbf{Rule} & \textbf{Rewrite and applicability} \\
    \hline

    \textbf{Commute} &
    $\operatorname{Op}(\mathit{Perm}_p(\mathbf{A}))
      \rightarrow
      \mathit{Perm}_p(\operatorname{Op}(\mathbf{A}))$
    \newline
    Requires uniform pointwise semantics in program logical coordinates.
    \\

    \textbf{Fusion} &
    $\mathit{Perm}_q(\mathit{Perm}_p(A))
      \rightarrow
      \mathit{Perm}_{q\circ p}(A)$
    \\

    \textbf{Identity} &
    $\mathit{Perm}_{\mathit{id}}(A)\rightarrow A$
    \\

    \textbf{Hoist} &
    $\mathit{Perm}_q(\mathit{Bcast}(A))
      \rightarrow
      \mathit{Bcast}(A)$
    \newline
    Requires $q$ to reorder only coordinates introduced by the broadcast.
    \\

    \hline
  \end{tabular}
\end{table}

All rewrites, including backend-specific rules, 
must preserve tensor-value semantics,
boundary axis mappings, and internal layout feasibility. For
\textbf{Hoist}, value preservation follows because values that differ only
in broadcast-introduced coordinates correspond to the same element of $A$,
regardless of whether those coordinates are realized temporally, spatially,
or in memory. The rewrite rules preserve boundary layout agreement and internal
feasibility, so cost-based extraction retains these properties.

In \Cref{fig:compile-example}(b), \textbf{Commute} moves the two
matching permutations through the pointwise addition.
\textbf{Fusion} and \textbf{Identity} then cancel the resulting
permutation with the inverse preceding \texttt{vround}.
These generic rules eliminate all three repairs, allowing
\texttt{vadd} to preserve the representation shared by
\texttt{vmpy} and \texttt{vround}.
Remaining repairs proceed to global layout solving.

\subsection{Global Layout Solving}
\label{sec:global-solving}

After repair simplification, \textbf{\textsc{Arknife}} encodes the
compatibility problem of \Cref{sec:layout-compatibility} as the binary
integer program in \Cref{alg:global-layout-solving}. Each
instruction-local axis already carries a hardware role declared by its
operand layout attributes, and the remaining unknown is which program root factor
plays that role.  Let $A_I$ be the instruction-local axes and $P$ the
atomic program root factors obtained by splitting program axes into
powers of two.  The variable $x_{a,p}$ indicates that $a\in A_I$ maps to
$p\in P$, thereby instantiating the embedding
$\phi_{o,t,b}(a)=p$ and inducing
$\Lambda_T(t)(p)=\Lambda_B(b)(a)$.  Constraints within an instruction
make $\phi$ a valid embedding, while constraints across uses of a shared
tensor encode $C_t\subseteq\Lambda_T(t)$. Thus, a feasible assignment
witnesses $\operatorname{Compat}$. Hard constraints enforce legality,
and soft preferences select among legal mappings.

Lines~2--5 collect the instruction-local factors, layout attributes, and
declared constraints. 
A non-broadcast factor selects exactly one program root factor,
$\sum_{p\in P}x_{a,p}=1$, whereas a broadcastable factor may remain
unmapped, $\sum_{p\in P}x_{a,p}\leq 1$.  For an
instruction-declared factor group $A$ with extent-compatible root
candidates $S\subseteq P$, the constraints forbid mappings outside $S$,
$x_{a,p}=0$ for $a\in A$ and $p\notin S$, and preserve injectivity
within the group,
$\sum_{a\in A}x_{a,p}\leq 1$ for $p\in S$.
Lines~6--8 merge the requirements induced by different uses of each
shared tensor.  If instruction-local factors $a$ and $a'$ denote the
same tensor root factor, their assignments must agree:
$x_{a,p}=x_{a',p}$ for every $p\in P$.  A fixed tensor root factor
$p_0$ instead pins the corresponding symbolic factor with
$x_{a,p_0}=1$.  An unmatched broadcast factor may remain unmapped.
These constraints ensure that the operand embeddings induce one
compatible tensor layout, as required by
$C_t\subseteq\Lambda_T(t)$.  The same loop records soft preferences for
contiguous mappings, suffix alignment, and reuse of compatible mappings
across instruction uses.
Lines~9--11 maximize the aggregate preference score subject to these constraints. 
Lines~12--13 solve the model, decode the selected
axis map, and apply it to the instruction bindings. The model contains
$|A_I||P|$ mapping variables. $|A_I|$ grows with
instruction and repair uses, while $|P|$ grows with the binary
decomposition of tiled program axes.

\begin{algorithm}[t]
\caption{Global axis-map solving}
\label{alg:global-layout-solving}
\small
\begin{algorithmic}[1]
\Require bound instruction graph $G_I$, program root factors $P$
\Ensure instruction-to-root embeddings and induced tensor layouts
\State $A_I\leftarrow\emptyset$; $\mathcal{L}\leftarrow\emptyset$;
       $\mathcal{C}\leftarrow\emptyset$; $\mathcal{O}\leftarrow\emptyset$
\For{each bound instruction use $i$}
  \State $(A_i,L_i)\leftarrow\mathrm{CollectAttrs}(i)$
  \State $A_I\leftarrow A_I\cup A_i$;
         $\mathcal{L}\leftarrow\mathcal{L}\cup L_i$
  \State $\mathcal{C}\leftarrow\mathcal{C}\cup\mathrm{DeclCons}(i,L_i)$
\EndFor
\For{each tensor $t$ with multiple layout attributes}
  \State $\mathcal{C}\leftarrow\mathcal{C}\cup\mathrm{MergeCons}(\mathcal{L}_t,P)$
  \State $\mathcal{O}\leftarrow\mathcal{O}\cup\mathrm{PrefTerms}(\mathcal{L}_t,P)$
\EndFor
\State $M\leftarrow\mathrm{BinaryILP}()$
\State $x\leftarrow M.\mathrm{BinaryVars}(A_I\times P)$;
        $M.\mathrm{addAll}(\mathcal{C}[x])$
\State $M.\mathrm{maximize}\bigl(\sum_{q\in\mathcal{O}}q[x]\bigr)$
\State $\sigma\leftarrow M.\mathrm{solve()}$;
        $\phi\leftarrow\mathrm{Decode}(\sigma,x)$
\State $G_I\leftarrow\mathrm{ApplyAxisMap}(G_I,\phi)$
\State \Return updated graph
\end{algorithmic}
\end{algorithm}

In \Cref{fig:compile-example}(b), shared-tensor constraints
align the vector roles across the instruction chain.
The solver selects $n=n_o\times n_v\times n_i$, mapping these
roles to $n_v$ and the two-way interleave roles of \texttt{vmpy}
and \texttt{vround} to $n_i$.
Since \texttt{vadd} declares no interleave role, $n_i$ remains
outside its tile. The remaining axes are $n_i\times n_o$ for
\texttt{vadd} and $n_o$ for the other instructions.
They may be realized through temporal iteration or hardware-axis
mappings, including thread or register axes.

\subsection{Layout-Preserving Lowering and Code Generation}
\label{sec:lowering}

Lowering refines the graph into an execution nesting while preserving
the instruction-boundary axis-to-hardware mappings fixed by $\Lambda$.

\begin{definition}[Layout-preserving transformation]
A graph rewrite $G\leadsto G'$ is \emph{layout-preserving} if, for every
instruction node $v\in V_I$, the boundary layout $\Lambda(v)$ computed in
\Cref{sec:global-solving} remains valid in $G'$. Specifically, every
$\operatorname{Legal}$ and $\operatorname{Compat}$ witness in $G$ carries over
to $G'$.
\end{definition}

Backends compose the following layout-preserving passes as needed.
\textbf{Spatial tiling} lowers a $\operatorname{SpatialOp}(p,\mathit{axes})$
with solved parallel mapping $\phi$ to a hardware-indexed loop over the
mapped root pieces, after a binary integer program fixes tile sizes under
hard per-unit capacity constraints at every program point.
\textbf{Temporal loop patching} wraps axes not yet covered by any loop in
$\operatorname{TemporalOp}$ without changing $\phi$.
\textbf{Memory planning} reuses storage across nonoverlapping tensor
lifetimes and assigns each temporary to the first sufficiently large free
block within its declared memory region.
\textbf{Code emission} traverses the refined graph, emitting loops from the
nesting structure and invoking each instruction's codegen callback with the
solved layout attributes. Repair nodes emit identically to compute
instructions.

For \Cref{fig:compile-example}(b), memory planning allocates
intermediate storage, and code emission lowers \texttt{vmpy},
\texttt{vadd}, and \texttt{vround} to Hexagon intrinsics
while preserving the solved mappings.
\section{Implementation}
\label{sec:implementation}

\textbf{\textsc{Arknife}} is implemented as a Python compiler framework. The frontend constructs an
operation graph whose tensors, axes, instruction bindings, repairs, layout
attributes, memory placements, and execution structure are updated by compiler
passes. Targets are also described in Python through two descriptions. A hardware description declares
parallel levels, memory spaces, capacities, and communication channels, while an
instruction description declares operand semantics, layout attributes, resource
effects, and code emission callbacks. The global layout and spatial tiling ILPs
are solved through PySCIPOpt 6.0.0, backed by SCIP 10.0.0. The implementation includes
x86, Qualcomm Hexagon, and NVIDIA GPU backends. 
Retargeting effort centers on per-target instruction descriptions.
Across our current backends, hardware hierarchy descriptions take
tens of lines, and each supported instruction fewer than ten.
All kernels on a target share these descriptions, independently
of kernel-specific scheduling decisions.
\section{Evaluation}
\label{sec:evaluation}

Our evaluation addresses three questions:

\noindent\textbf{1.} How does \textbf{\textsc{Arknife}} perform against vendor libraries and
retargetable compilers across our three backends?
(\Cref{sec:runtime-performance})

\noindent\textbf{2.} How effectively does \textbf{\textsc{Arknife}} reduce
the runtime overhead of layout handling compared with prior retargetable
compilation?
(\Cref{sec:layout-handling-breakdown})

\noindent\textbf{3.} What compilation overhead does global layout solving
introduce, and how does it scale with tensor size and graph structure?
(\Cref{sec:layout-solving-overhead})

\subsection{Experimental Setup}

We evaluate \textbf{\textsc{Arknife}} on three backends including x86 CPU, 
Qualcomm Hexagon NPU, and NVIDIA A100. x86 experiments
run on a dual-socket Intel Xeon Gold 6348 (112 threads, AVX-512/VNNI).
We compare against oneDNN 3.7.1~\cite{onednn}, Halide~\cite{halide2013},
Hydride~\cite{hydride2024}, and MISAAL~\cite{misaal2025}. Compilation-overhead
measurements (\Cref{sec:layout-solving-overhead}) were collected on the
same machine, running Ubuntu 22.04.4 LTS with 2 TiB of memory. 
The oneDNN comparison runs both systems with 16 threads and
measures steady-state per-call latency with weights prepared on both
sides. Halide uses the schedules distributed with Hydride without retuning.
Hexagon experiments use Hexagon SDK 6.0.0.0 with Hexagon tools 8.7.08 for the
compiler-baseline comparison. The vendor baseline uses QAIRT with
QNN 2.28.0~\cite{qnn} on a Snapdragon SM8650 device. 
On an NVIDIA A100 80GB PCIe GPU, we compare FP16
MHA against PyTorch~\cite{pytorch2019}, Triton~\cite{triton2019},
FlashAttention~\cite{flashattention2022}, and TileLang~\cite{tilelang2025}, and
TF32 GEMM against cuBLAS~\cite{cublas}, Triton, and TileLang. 
For FP16 MHA, our Triton baseline is a forward-only adaptation of the
official Triton 3.2 fused-attention tutorial and autotunes configurations
for each sequence length and head dimension. The TileLang baseline
specializes its official v0.1.13 BSHD example using the published
$128\times128$ tile configuration, while FlashAttention uses the official
library implementation.

The evaluated targets impose distinct operand layouts including x86 VNNI packing,
Hexagon HVX/HMX lane layouts and VTCM tiles, and A100 warp/register fragment
mappings for the \texttt{ldmatrix}--MMA path.

\subsection{Runtime Performance Across Backends}
\label{sec:runtime-performance}

\subsubsection{x86}

\begin{figure}[t]
  \centering
  \includegraphics[width=\linewidth]{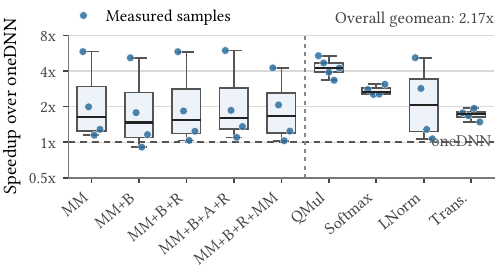}
  \caption{x86 speedup over oneDNN across input shapes (log scale).
  Boxes show the interquartile range, center lines the median,
  whiskers the minimum and maximum, and dots individual shapes.
  MM, B, A, and R denote MatMul, Bias, Add, and ReLU.}
  \Description{Box plots with individual-shape data points for nine
  x86 workloads. A dashed line marks the oneDNN baseline at one times.
  The overall geometric-mean speedup is 2.17 times.}
  \label{fig:onednn-speedup}
\end{figure}

\textbf{\textsc{Arknife}} achieves a $2.17\times$ geomean speedup over
oneDNN across nine x86 workloads (\Cref{fig:onednn-speedup}),
$1.91\times$ on matrix and $2.55\times$ on scalar workloads. The
matrix gain is largest at small shapes, where oneDNN's larger
per-call overhead weighs more, and closes to parity at the largest
shape. Both systems target AVX-512 VNNI. \textbf{\textsc{Arknife}} binds MatMul to
\texttt{vpdpbusd}, packing $K$ into groups of four within each
32-bit lane with an i32 accumulator per lane, and global layout
solving (\Cref{sec:global-solving,sec:lowering}) keeps this layout
through Bias, Add, ReLU, and the second MatMul without boundary
conversions. Its programming interface hides this layout reasoning
from users.

\begin{figure*}[t]
  \centering
  \begin{subfigure}[t]{0.49\textwidth}
    \centering
    \includegraphics[width=\linewidth]{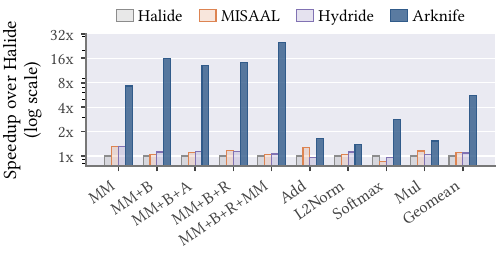}
    \caption{x86 CPU.}
    \label{fig:retargetable-x86}
  \end{subfigure}\hfill
  \begin{subfigure}[t]{0.49\textwidth}
    \centering
    \includegraphics[width=\linewidth]{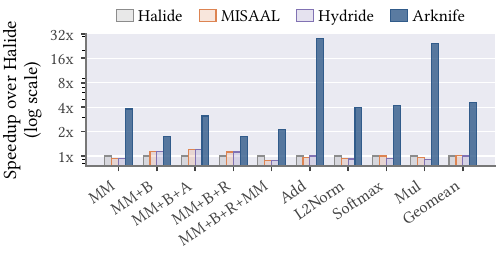}
    \caption{Qualcomm Hexagon NPU.}
    \label{fig:retargetable-hexagon}
  \end{subfigure}
  \caption{Per-workload performance normalized to Halide on x86 and Hexagon (higher is
  better, log scale). MM, B, A, and R denote MatMul, Bias, Add, and ReLU.}
  \Description{Two vertical grouped bar charts comparing Halide, MISAAL,
  Hydride, and \textbf{\textsc{Arknife}} across all measured x86 and Hexagon
  workloads, plus a geomean group. \textbf{\textsc{Arknife}} is generally
  higher than the other baselines, with the largest gains on several matrix
  and elementwise workloads.}
  \label{fig:retargetable-geomean}
\end{figure*}

Across the nine x86 workloads in \Cref{fig:retargetable-x86},
\textbf{\textsc{Arknife}} achieves $5.08\times$ and $5.13\times$ geometric
mean speedups over MISAAL and Hydride, respectively. The per-workload
gaps grow with the number of fused boundaries. MatMul alone reaches
$7.28\times$ over Halide, the fused variants with one or two
epilogues reach $12.98\times$ to $15.82\times$, and
MatMul+Bias+ReLU+MatMul reaches $25.35\times$, while both baselines
remain below $1.35\times$ throughout. Per-operator schedules rebuild
the solved accumulator layout at every epilogue and materialize and
reorganize the intermediate for the second MatMul.
\Cref{sec:layout-handling-breakdown} examines these boundaries in
more depth, quantifying the layout handling of both strategies.

\subsubsection{Hexagon}

Across the nine Hexagon workloads in \Cref{fig:retargetable-hexagon},
\textbf{\textsc{Arknife}} achieves $4.53\times$ and $4.58\times$ geometric
mean speedups over MISAAL and Hydride, respectively. Unlike x86,
the largest gaps occur on quantized Add and Mul, at $28.29\times$ and
$24.10\times$ over Halide, where one vector op per element makes
lane organization the bottleneck.
On the matrix chains the per-boundary tax is smaller. Pointwise
epilogues are cheap to fuse on HVX, and unlike on x86, where
\textbf{\textsc{Arknife}} lowers the whole chain in one kernel, the
evaluated MatMul+Bias+ReLU+MatMul is compiled as two
\textbf{\textsc{Arknife}} kernels around a materialized global
intermediate, leaving no cross-boundary layout win to recover.
\Cref{sec:layout-handling-breakdown} quantifies the Add and Mul
contrast through dynamic layout moves and conversions.

\begin{figure}[t]
  \centering
  \includegraphics[width=\linewidth]{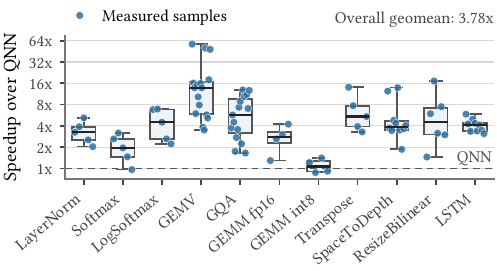}
  \caption{Hexagon speedup over QNN across operator configurations
  (log scale). Boxes show the interquartile range, center lines
  the median, whiskers the minimum and maximum, and dots individual
  configurations.}
  \Description{Box plots with individual-configuration data points
  for eleven Hexagon operators. A dashed line marks the QNN baseline
  at one times. The overall geometric-mean speedup is 3.78 times.}
  \label{fig:qnn-speedup}
\end{figure}

\textbf{\textsc{Arknife}} achieves a $3.78\times$ geomean speedup over QNN
across eleven Hexagon operators (\Cref{fig:qnn-speedup}).
Transpose ($5.96\times$) instantiates the permutation-repair machinery of
\Cref{sec:repair}. Global layout solving records the axis
correspondence of each $64\times64$ FP16 tile, and layout-preserving
lowering realizes the permutation with vector loads, an in-register
\texttt{vshuff} network, and vector stores without scalarizing it.
On Int8 GEMM, where both sides target the HMX matrix unit and its
instruction-visible operand layout (\Cref{fig:retargetable-gemm}(d)),
\textbf{\textsc{Arknife}} is on par with the mature vendor path
($1.08\times$) on the tested configurations, and the per-operator gains
remain workload-dependent.

\noindent\textbf{\textit{End-to-end evaluation.}}
For single-batch, single-step decode based on Qwen3-0.6B without
additional cross-operator fusion, \textbf{\textsc{Arknife}} reaches
$0.87\times$, $1.07\times$, $2.34\times$, and $2.77\times$ the QNN
throughput as the sequence grows from 1024 to 8192. The speedup grows
with context because execution shifts from projection, whose weight
streaming is bandwidth-bound and layout-insensitive, to GQA decode,
where layout dominates as each KV head is reused by multiple query
heads. The per-operator comparison gives both systems raw
weights with no offline conversion, isolating mapping quality. The
end-to-end protocol instead follows deployment, where each side
prepares its weights once and QNN serves them from the compiled graph
rather than uploading them per inference. Automatic weight
preprocessing, weight residency, and megakernel
fusion are left for future work.

\subsubsection{NVIDIA A100}

\begin{figure*}[t]
  \centering
  \begin{subfigure}[t]{0.49\textwidth}
    \centering
    \includegraphics[width=\linewidth]{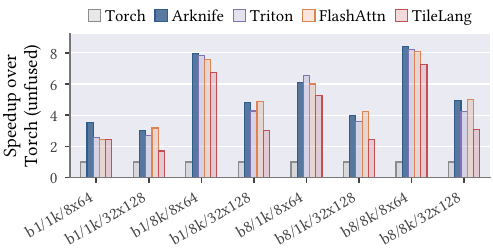}
    \caption{FP16 MHA}
    \label{fig:gpu-mha}
  \end{subfigure}\hfill
  \begin{subfigure}[t]{0.49\textwidth}
    \centering
    \includegraphics[width=\linewidth]{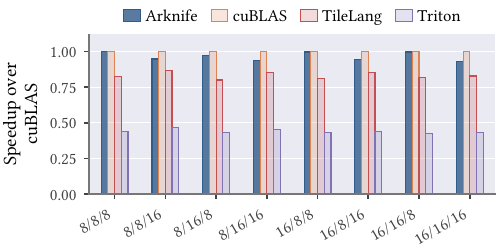}
    \caption{TF32 GEMM}
    \label{fig:gpu-gemm}
  \end{subfigure}
  \caption{Evaluation on A100.
  MHA reports speedup over unfused PyTorch;
  labels are batch/sequence length/heads$\times$head dimension.
  GEMM reports speedup over cuBLAS;
  labels are $M/N/K$ in thousands.}
  \Description{Two grouped bar charts. The first compares MHA speedup
  over unfused PyTorch across eight configurations. The second compares
  TF32 GEMM speedup over cuBLAS across eight matrix sizes, computed as
  cuBLAS latency divided by each implementation's latency.}
  \label{fig:gpu-reference}
\end{figure*}

Across eight FP16 MHA configurations (\Cref{fig:gpu-reference}),
\textbf{\textsc{Arknife}} achieves a $5.04\times$ geometric mean
speedup over unfused PyTorch eager, which executes attention through
separate kernel launches. The corresponding speedups over the
same baseline are $4.56\times$ for Triton, $4.83\times$ for FlashAttention, and $3.53\times$ for TileLang. 
On TF32 GEMM,
\textbf{\textsc{Arknife}} reaches 97\% of cuBLAS performance,
and outperforms Triton and TileLang by $2.20\times$ and $1.16\times$,
respectively.

\begin{figure}[t]
  \centering
  \includegraphics[width=\linewidth]
  {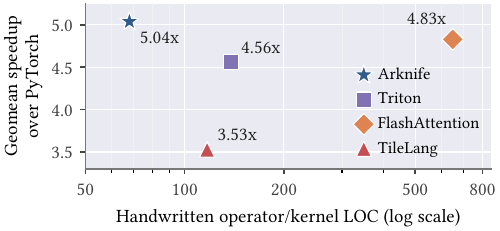}
  \caption{FP16 MHA geometric-mean speedup over unfused PyTorch versus
  handwritten LOC on A100.}
  \Description{A scatter plot comparing geometric-mean MHA speedup over
  PyTorch with handwritten operator or kernel LOC for
  \textbf{\textsc{Arknife}}, Triton, FlashAttention, and TileLang.
  \textbf{\textsc{Arknife}} has the highest speedup and the lowest LOC among
  the four fused implementations.}
  \label{fig:gpu-mha-loc}
\end{figure}

For MHA, fusing both matrix products and online softmax into one tiled
kernel avoids attention-matrix materialization and intermediate launches,
providing the primary gain over unfused PyTorch.
\textbf{\textsc{Arknife}} uses swizzled shared-memory tiles, asynchronous
copies, \texttt{ldmatrix}, and MMA, satisfying tensor-core fragment
layout requirements without manual specification.
\Cref{fig:gpu-mha-loc} uses handwritten operator and kernel LOC as a coarse
proxy for programming complexity, excluding blank and comment lines, shared
infrastructure, and benchmark code. Among these implementations,
\textbf{\textsc{Arknife}} reaches the top geometric-mean speedup
using only 68 lines of code, versus 117--650 for the others.
Logical tensor axes and reusable instruction descriptions let the
compiler derive lane and register mappings, leaving hardware
mechanics out of each kernel.
For TF32 GEMM, \textbf{\textsc{Arknife}} combines multistage pipelining with
layout-compatible transfers from shared-memory tiles to MMA operands. 
The highly optimized cuBLAS implementation leaves limited
room for further gains.

\subsection{Layout-Handling Breakdown}
\label{sec:layout-handling-breakdown}

To investigate the observed speedups, we compare dynamic
layout-handling instruction counts between \textbf{\textsc{Arknife}}
and Hydride in four settings:
quantized x86 MatMul + Bias + ReLU (Setting I), the same graph with an
additional consuming MatMul (Setting II), and Hexagon quantized Add
(Setting III) and Mul (Setting IV). Settings I and II use $B=1$ and
$M=N=K=256$, while Settings III and IV use $M=N=1024$.

\emph{Move} transfers tensor values between memory or register spaces, while
\emph{Convert} changes lane, axis, tile, parallel-distribution, or element
representation. Instructions combining memory access and
conversion count as Convert. \emph{Scalarized} is the fraction of layout
instructions that touch one element per execution (e.g., \texttt{vpinsrw},
byte-granular memory access, scalar shifts) rather than a whole vector.
Operator arithmetic, address
generation, control flow, and ABI bookkeeping are excluded from these
categories but remain in the total instruction count. Dynamic counts are
derived from basic-block execution frequencies on x86 and simulator PC
traces on Hexagon.
\Cref{tab:layout-handling-breakdown} reports rounded counts. Layout is Move
plus Convert, Share is its fraction of all dynamically executed instructions,
and H/A is the Hydride-to-\textbf{\textsc{Arknife}} ratio in Layout.

\begin{table}[t]
  \centering
  \caption{Dynamic layout handling for the four evaluation settings.}
  \Description{Rounded dynamic Move, Convert, and Scalarized
  layout-handling instruction counts for Arknife, with and without
  repair-graph optimization, and for Hydride, on two x86 fused graphs
  and Hexagon quantized Add and Mul.}
  \label{tab:layout-handling-breakdown}
  \small
  \setlength{\tabcolsep}{2.5pt}
  \begin{tabular}{@{}llrrrr@{}}
    \toprule
    Setting & System & Move/Convert & Layout (Share) & Scalarized & H/A \\
    \midrule
    \multirow{3}{*}{I}
      & \textbf{\textsc{Arknife}}
      & 407K / 139K & 546K (18.3\%) & 0.0\% & -- \\
      & \quad w/o e-graph
      & 407K / 139K & 546K (18.3\%) & 0.0\% & -- \\
      & Hydride
      & 32.8K / 9.30M & 9.33M (37.0\%) & 91.3\% & \textbf{17.1$\times$} \\
    \cmidrule(l){1-6}
    \multirow{3}{*}{II}
      & \textbf{\textsc{Arknife}}
      & 467K / 664K & 1.13M (16.5\%) & 0.0\% & -- \\
      & \quad w/o e-graph
      & 467K / 664K & 1.13M (16.5\%) & 0.0\% & -- \\
      & Hydride
      & 12.7M / 45.3M & 58.1M (43.8\%) & 74.2\% & \textbf{51.4$\times$} \\
    \cmidrule(l){1-6}
    \multirow{3}{*}{III}
      & \textbf{\textsc{Arknife}}
      & 67.6K / 31.7K & 99.3K (27.4\%) & 0.0\% & -- \\
      & \quad w/o e-graph
      & 135K / 80.9K & 216K (44.4\%) & 0.0\% & -- \\
      & Hydride
      & 4.03M / 4.49M & 8.52M (90.7\%) & 73.8\% & \textbf{85.8$\times$} \\
    \cmidrule(l){1-6}
    \multirow{3}{*}{IV}
      & \textbf{\textsc{Arknife}}
      & 69.6K / 30.7K & 100K (26.9\%) & 0.0\% & -- \\
      & \quad w/o e-graph
      & 81.9K / 63.5K & 145K (34.4\%) & 0.0\% & -- \\
      & Hydride
      & 3.50M / 3.72M & 7.22M (93.6\%) & 80.7\% & \textbf{71.9$\times$} \\
    \bottomrule
  \end{tabular}
\end{table}

On x86, Hydride rebuilds vectors lane by lane at each fused epilogue and
materializes and reorganizes the intermediate for the second MatMul.
79--81\% of its Convert instructions are \texttt{vpinsrw} inserts, 91.3\%
(I) and 74.2\% (II) of its layout handling is scalarized, and
\textbf{\textsc{Arknife}} reduces layout handling by $17.1\times$ and
$51.4\times$. On Hexagon, byte-granular payload moves and scalar shifts
compose 73.8\% (Add) and 80.7\% (Mul) of Hydride's layout handling,
reaching 67.0\% (Add) and 75.6\% (Mul) of all dynamic instructions, while
every layout instruction counted for \textbf{\textsc{Arknife}} moves a
whole vector, reducing layout handling by $85.8\times$ and $71.9\times$.

\noindent\textbf{\textit{Ablating repair-graph optimization.}}
Disabling the e-graph pass leaves the x86 pipelines unchanged, 
as GCC -O2 already eliminates the register copies
and pack/unpack round trips it could remove.
On Hexagon the pass absorbs the interleavings that widening multiplies
(e.g., \texttt{vmpy}, \texttt{vmpye}/\texttt{vmpyoacc}) natively produce.
Hoisting one-sided permutes across lane-wise operations cancels them
against their inverses, so the \texttt{vshuff}/\texttt{vdeal} repairs of
provisional binding are composed away instead of materializing, cutting
layout handling by $2.18\times$ (III) and $1.45\times$ (IV).

\subsection{Global Layout-Solving Overhead}
\label{sec:layout-solving-overhead}

\begin{table}[t]
  \centering
  \caption{Global layout solving overhead.}
  \Description{A table reporting axis-map pass and ILP solve time on
  representative x86 and Hexagon workloads.}
  \label{tab:solver-overhead}
  \small
  \begin{tabular}{@{}llcc@{}}
    \toprule
    Backend & Workload & Axis-map (ms) & ILP solve (ms) \\
    \midrule
    x86 & MatMul & 177.3 & 19.6 \\
    x86 & MatMul + B + ReLU & 340.6 & 34.9 \\
    Hexagon & HMX GEMM & 228.5 & 44.7 \\
    Hexagon & Decode & 1862.1 & 296.1 \\
    Hexagon & Transpose & 56.1 & 6.5 \\
    \bottomrule
  \end{tabular}
\end{table}

We measure offline layout-solving cost by instrumenting the axis-map
pass and SCIP calls. Reported times sum all invocations during
instruction binding and layout transformation optimization and on the
final bound graph. \Cref{tab:solver-overhead} reports medians over five
runs. Axis-map includes constraint construction, ILP solving, and
result application, and ILP solve includes only SCIP optimization. 
The measured settings use $M=N=K=256$ for the x86 workloads,
$M=N=K=1024$ for HMX GEMM, $1024\times1024$ for Transpose, and
$S=4096$, $d_{\mathrm{model}}=4096$, and $H_q=H_{kv}=1$ for GQA Decode. 
Total axis-map time ranges from 56.1 ms to 1.86 s.
SCIP accounts for only 10--20\% of this time, so constraint
construction and result application dominate. Decode incurs the
highest cost due to more axis-map calls (nine versus five for x86
MatMul) and more instruction and repair constraints per call.

To evaluate sensitivity to problem size, we sweep x86 MatMul from $256^3$ to
$1024^3$ and GQA Decode from sequence length 1024 to 8192, while keeping
$d_{\mathrm{model}}=4096$ and $H_q=H_{kv}=1$. The median ILP solve time remains
within 15.9--19.1 ms for MatMul and 273.2--291.7 ms for Decode. 
With the binary axis representation, doubling
a dimension adds at most one root piece, so solve time is more sensitive
to graph constraints than to tensor size.

\section{Related Work}
\label{sec:related-work}

\noindent\textbf{\textit{Tensor DSLs and scheduling systems.}}
Schedule-separation systems~\cite{halide2013,tvm2018} decouple algorithms
from optimization schedules, while automatic scheduling
systems~\cite{ansor2020,metaschedule2022} search schedules over tensor IRs.
Tiled programming systems~\cite{triton2019,tensorir2022,tilelang2025,
tirx2026} expose layout and memory placement in the program and support
cross-backend code generation.
Polyhedral and schedule-search techniques further extend this
design space~\cite{tensor_comprehensions2018,tiramisu2018,
flextensor2020,dietcode2022,sparsetir2022,xla2020}.
TensorIR~\cite{tensorir2022} checks intrinsic layout constraints
during scheduling, TileLang~\cite{tilelang2025} propagates layouts
by operator priority, and TIRx~\cite{tirx2026} dispatches primitives
from explicit storage layouts.
\textbf{\textsc{Arknife}} instead reconciles target-declared operand
contracts across instruction boundaries.

\noindent\textbf{\textit{Retargetable instruction selection.}}
Semantics-driven systems~\cite{vegen2021,diospyros2021,lift2017,amos2022}
use formal instruction descriptions to retarget computation across hardware. 
UNIT~\cite{unit2021} detects tensorized-instruction opportunities and
rewrites loop and data organizations to match instruction templates
across architectures.
Hydride and MISAAL~\cite{hydride2024,misaal2025} apply semantics-driven
retargeting across the full compilation pipeline. 
Equality saturation~\cite{tensat2021,rake2022,pet2021} and
hardware-exposed scheduling~\cite{exo2022,fireiron2020,isaria2024}
push retargeting further but still leave instruction-level layout to the
programmer or backend.
These systems primarily establish or exploit computational equivalence
for instruction selection and rewriting.  \textbf{\textsc{Arknife}}
adds explicit layout requirements for memory placement, axes organization,
and parallel distribution, checks their agreement at shared tensor
boundaries, and inserts target-provided repairs when the selected
instructions disagree.

\noindent\textbf{\textit{Libraries, graph compilers, and layout optimization.}}
Vendor libraries~\cite{onednn,cublas,cutlass,qnn} encode layouts in
target-specific, hand-tuned implementations. 
Graph compilers~\cite{glow2018,relay2019,taso2019} represent layout as an
operator-level attribute rather than a per-instruction constraint. 
Layout-aware kernel systems~\cite{roller2022,welder2023,
flashattention2022,flashattention22023,vllm2023} and mobile NPU
work~\cite{hao2026scaling} demonstrate that explicit layout control is
essential for performance. Operator
fusion~\cite{dnnfusion2021,apollo2022,rammer2020}, memory-coordinated
optimization~\cite{checkmate2020,unity2022,magis2024,trinity2026},
infrastructure-level retargeting~\cite{mlir2021,nncase2025} 
address complementary aspects.
Linear Layouts~\cite{linear_layouts2026} unifies GPU layout
representations and conversion code generation through algebra
over $\mathbb{F}_2$, with heuristic layout propagation in Triton.
Tensor Seeks Layout~\cite{tensor_seeks_layout2026} formulates global
layout selection over an operator graph, minimizing operator and
conversion costs before fusion and tiling.
Mirage~\cite{mirage2025} searches layout-level rewrites through multi-level
superoptimization for single-GPU inference.
\textbf{\textsc{Arknife}} instead inserts and simplifies
target-provided repairs on the bound instruction graph and then solves
layout globally.

\section{Conclusion}
\label{sec:conclusion}

We present \textbf{\textsc{Arknife}}, a retargetable compiler that treats
instruction-visible layout requirements as first-class instruction
semantics and globally coordinates layout assignments and repairs. It 
achieves $5.08\times$ and $4.53\times$ geometric mean speedups over MISAAL
on x86 and Hexagon, respectively, while reaching 97\% of cuBLAS performance
on A100 TF32 GEMM. These results show that explicit layout requirements 
can combine retargetability with efficient data paths across diverse
accelerators.

\bibliographystyle{ACM-Reference-Format}
\bibliography{refs}
\end{document}